\documentclass[%
reprint,  amsmath,amssymb,
aps,prl
]{revtex4-2}

\usepackage{graphicx}
\usepackage{systeme,braket,color}
\usepackage[colorlinks=true,linkcolor=blue]{hyperref}

\begin{document}
	
	\preprint{APS/123-QED}
	
	\title{Non-reciprocal waves in two-dimensional electron systems with temperature gradient}
	
	\author{Aleksandr S. Petrov}
	\email{petrov.as@mipt.ru}

	\affiliation{Laboratory of 2D Materials for Optoelectronics, Moscow Institute of Physics and Technology, Dolgoprudny, Russia}

	\date{April 07, 2025}
	
	\begin{abstract}
		
		 We demonstrate that the interaction of dc temperature gradient with ac magnetic field in temperature-biased two-dimensional electron systems leads to formation of a new electromagnetic mode, a two-dimensional thermomagnetic wave (2d TMW). This wave is transverse electric and non-reciprocal, and its damping rate can be lower than that of conventional 2d plasma waves. The Q-factor of 2d TMW is independent of the wave vector. Numerical estimates show that in state-of-the-art two-dimensional electrons systems the 2d TMW Q-factor is the order of $10^{-3}$. We discuss possible ways to overcome this issue.
		
	\end{abstract}
	
	\maketitle
	
\section{Introduction}
	
Many applications require the transmission of a time-modulated signal in a given direction without reflections. In electronics, the simplest way to achieve this goal relies on matching the circuit loads (and when the signal reaches the end of the circuit, it is completely absorbed by the receiver). In optics common approaches include the use of optical isolators and circulators~\cite{jalas2013and} based on time-dependent~\cite{yu2009complete}, nonlinear~\cite{tocci1995thin,poulton2012design} or non-reciprocal media~\cite{shoji2012mzi,petrov2024high}.

The most common example of non-reciprocity in optics arises for ordinary materials biased by external magnetic field (gyrotropic media). Less common options include biasing conducting media (solid state electronic plasma) by electric field (which leads to dc current)~\cite{mikhailov1998plasma}, temperature gradient~\cite{brown2001thermal} or intrinsic material non-reciprocity (topological effects)~\cite{berry1984quantal,xiao2010berry,petrov2021plasmonic}. In these cases the external bias leads to formation of unidirectional electromagnetic modes with propagation direction uniquely defined by the bias direction. 

The usual drawback of using (semi)conducting materials is the relatively high Ohmic loss. Still, if the restoring force acting on charge cariers has magnetic origin, the Ohmic loss is greatly diminished. Such oscillations include galvanomagnetic and thermomagnetic waves (GMWs and TMWs, respectively) in three-dimensional electron systems (3DESs). These waves were predicted by Gurevich and Gel'mont~\cite{gurevich1963thermomagnetic,gurevich1967nonlinear} in 1960s and observed by Kopylov in late 1970s in Bi monocrystals~\cite{kopylov1978thermomagnetic}. For these waves, the restoring force arises as a result of rather unusual interaction between the \textit{ac magnetic field of the wave} with external bias (charge carrier drift for GMWs and temperature gradient for TMWs). Interestingly, these waves are well-defined even at frequencies order of 100 Hz (whereas the typical charge carrier momentum relaxation rate $1/\tau_p \simeq 1$\,THz).

With the widespread implementation of 2D materials it seems natural to search for the analogs of 3D GMWs and TMWs in two dimensions. Recently we conducted the corresponding analysis for galvanomagnetic waves and found that 2D GMWs (a) exist (b) are unidirectional (c) their damping rate can be lower than the standard $1/2\tau_p$ plasmon damping rate (d) the 2D GMW Q-factor does not exceed the 3D GMW Q-factor~\cite{petrov2024high}. Still, the proper design of 2DES surroundings may help to increase 2D TMW Q-factor. 

The 3D TMW is isothermal and has the following dispersion~\cite{gurevich1967nonlinear}:
\begin{equation}\label{eq-disp-3DTMW}
	\omega = -\alpha_1 c (\mathbf k,\nabla T) - i\frac{k^2 c^2}{4\pi\sigma_\tau^{3d}},
\end{equation}
where $\alpha_1$ is the Nernst-Ettingshausen coefficient, $\mathbf k$ is the wave vector, $\nabla T$ is the constant temperature gradient in the 3DES, $c$ is the speed of light, $\sigma_\tau^{3d} = e^2 n_{3d}\tau_p/m$ is the static 3DES conductivity, $e>0$ is the elementary charge, $n_{3d}$ is the background carrier density, $\tau_p$ is the effective carrier momentum relaxation time, $m$ is the effective mass. Eq.(\ref{eq-disp-3DTMW}) dictates that the TMW is unidirectional, and its quality factor $\mathrm{Re}\,\omega/\mathrm{Im}\,\omega \to\infty$ in the long wavelength limit.

The aim of this work is to search for the two-dimensional analog of 3D thermomagnetic wave~\cite{gurevich1967nonlinear} and analyze its properties.

\section{Theoretical model}

\begin{figure}
	\centering
	\includegraphics[width=\linewidth]{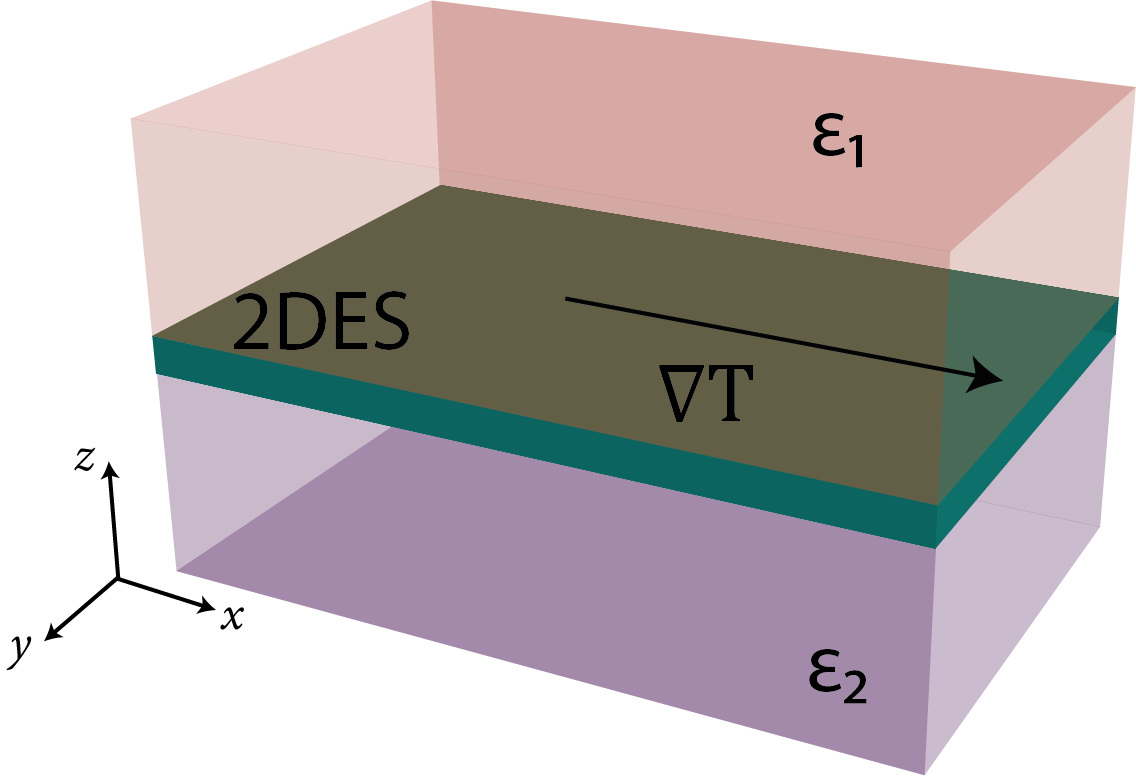}
	\caption{Schematic view of a host system for two-dimensional TMW. A 2DES with homogeneous carrier density is subject to temperature gradient in $x$-direction, which leads to formation of compensating electric field. No charge carrier drift is present. The structure is sandwiched between dielectrics with permittivities $\varepsilon_1,\varepsilon_2$. The main part of the paper deals with the case $\varepsilon_1=\varepsilon_2$ for simplicity.}
	\label{Fig-setup-scheme}
\end{figure}

We consider an infinite two-dimensional electron system (2DES) with homogeneous carrier density $n_0$ that is sandwiched between two materials with permittivities $\varepsilon_i, i =1,2$. A constant temperature gradient $\nabla T$ is applied along the $x$-axis, which leads to formation of a compensating electric field; no current is present in the system (Fig.~\ref{Fig-setup-scheme}).

The general material relation for an electric field $\mathbf{E}$ that arises in isotropic conducting system subject to magnetic field $\mathbf{H}$, electric current $\mathbf{j}$ and temperature gradient $\nabla T$ reads:
\begin{equation}\label{eq-transport-coef}
	\begin{split}
	\mathbf{E} = &\rho_0\mathbf{j} + \rho_1 \mathbf j\times\mathbf H + \rho_2\mathbf H(\mathbf j,\mathbf H) + \\ &\alpha_0\nabla T + \alpha_1 \nabla T\times \mathbf H + \alpha_2\mathbf H(\nabla T,\mathbf H),
	\end{split}
\end{equation}
where $\rho_0 - \rho_2$ is the resistance, $\rho_1$ is the Hall resistance, $\alpha_0$ is the thermoelectric coefficient, $\alpha_1$ is the Nernst-Ettingshausen coefficient, $\rho_2$ and $\alpha_2$ are longitudinal magnetothermoelectric coefficients.

Assume that an isothermal wave $\delta\chi e^{i\mathbf{k}\mathbf{r}-|k_z z| - i\omega t}$ is localized near 2DES sandwiched between two dielectrics (see Fig.~\ref{Fig-setup-scheme}) and its magnitude $\delta\chi$ is small ($\delta\chi$ can be any quantity associated with the wave). Here $\mathbf k$ is the wave vector in the 2DES plane, $k_z$ is responsible for wave localization near 2DES plane $z=0$, $\omega$ is the wave frequency. Linearizing Eq.(\ref{eq-transport-coef}) we arrive at
\begin{equation}
	\delta\mathbf{E} = \rho_0\delta\mathbf{j} + \alpha_1 \nabla T\times \delta\mathbf H,
\end{equation}
With the help of the Faraday's law the above equation can be cast into the form
\begin{equation}
	\delta\mathbf{j} = \hat{\sigma}^{TM}\delta\mathbf{E},
\end{equation}
where the effective "thermomagnetic" conductivity $\hat{\sigma}$ is given by
\begin{equation}
	\hat{\sigma}^{TM} = \sigma_0
	\begin{pmatrix}
		1 & 0 \\
		\Omega/\omega\cdot \tan\alpha & 1-\Omega/\omega
	\end{pmatrix},
\end{equation}
where $\Omega = -\alpha_1 c (\mathbf k,\nabla T)$ is the a typical TMW frequency (compare with Eq.~(\ref{eq-disp-3DTMW})). The derived conductivity tensor is asymmetric, which is a natural consequence of a selected direction in the system (parallel to $\nabla T$).

Now our goal is to search for eigen modes in the scheme of Fig.~\ref{Fig-setup-scheme} with 2DES conductivity $\hat{\sigma}^{TM}$. The derivation closely follow the derivation of 2D GMW spectrum~\cite{petrov2024high}. We define area $z>0$ be area $\mathrm{I}$ and area $z<0$ be area $\mathrm{II}$ and let $\varepsilon_1 = \varepsilon_2 = \varepsilon$ for simplicity (the asymmetric case $\varepsilon_1\ne\varepsilon_2$ does not qualitatively affect the results, see the Discussion section). Then we search for eigenmodes in the form of linear combination of TE and TM waves:
\begin{gather}
	\label{eq-polarization-TE}
	\mathbf{E}_{\mathrm{I}} = \left[\mathrm{TE}_{\mathrm{I}}
	\begin{pmatrix}
		-\sin\alpha \\
		\cos\alpha\\
		0
	\end{pmatrix}
	+
	\mathrm{TM}_{\mathrm{I}}
	\begin{pmatrix}
		-ik_z\cos\alpha/k \\
		-ik_z\sin\alpha/k \\
		1
	\end{pmatrix}
	\right]
	\mathcal{E}_{\mathrm{I}}(\mathbf{r},t); 
	\\
	\label{eq-polarization-TM}
	\mathbf{E}_{\mathrm{II}} = \left[\mathrm{TE}_{\mathrm{I}}
	\begin{pmatrix}
		-\sin\alpha \\
		\cos\alpha\\
		0
	\end{pmatrix}
	+
	\mathrm{TM}_{\mathrm{II}}
	\begin{pmatrix}
		ik_z\cos\alpha/k \\
		ik_z\sin\alpha/k \\
		1
	\end{pmatrix}
	\right]
	\mathcal{E}_{\mathrm{II}}(\mathbf{r},t), 
\end{gather}
where $\mathrm{TE}_{\mathrm{I, II}}$ and $\mathrm{TM}_{\mathrm{I, II}}$ are the amplitudes of TE and TM electric field in the  corresponding areas,
\begin{equation}
\mathcal{E}_{\mathrm{I,II}}(\mathbf{r},t) = \exp\left(ik\cos\alpha x+ik\sin\alpha y \mp k_z z-i\omega t\right),
\end{equation}
$\alpha$ is the angle between the wave vector and the thermal gradient direction, $k_z = \sqrt{k^2-\varepsilon k_0^2}$, and $k_0 = \omega/c$.

Then we evaluate the corresponding magnetic fields via the Faraday's law and apply the boundary conditions on the tangential components of the electric and magnetic fields. As a result, we arrive at a linear system which acquires diagonal form if we change variables to $\mathrm{TE}_\pm =1/2 (\mathrm{TE}_{\mathrm{I}}\pm\mathrm{TE}_{\mathrm{II}}) $ and $\mathrm{TM}_\pm =1/2 (\mathrm{TM}_{\mathrm{I}}\pm\mathrm{TM}_{\mathrm{II}}) $:
\begin{equation}
	\begin{pmatrix}
		\hat{M}_1 & 0 \\
		0 & \hat{M}_2  \\
	\end{pmatrix}
	\begin{pmatrix}
	\left\{\mathrm{TE}_-,\mathrm{TM}_+\right\}^T \\
	\left\{\mathrm{TE}_+,\mathrm{TM}_-\right\}^T  \\
	\end{pmatrix}
	= 0,
\end{equation}
where the upper index $T$ denotes transposition operation,
\begin{equation}
	\hat{M}_1 = 
	\begin{pmatrix}
		-\sin\alpha & -i\kappa_z\cos\alpha/\kappa \\
		\cos\alpha & -i\kappa_z\sin\alpha/\kappa
	\end{pmatrix},
\end{equation}
$\kappa_z = k_z/k_0$, $\kappa = k/k_0$,
\begin{widetext}
\begin{equation}
	\hat{M}_2 = 
		\begin{pmatrix}
		-i\sin\alpha\Sigma_{xx} + i\cos\alpha\Sigma_{xy} + \kappa_z\sin\alpha & \cos\alpha\Sigma_{xx} \kappa_z/\kappa + \sin\alpha\Sigma_{xy} \kappa_z/\kappa - i\sqrt{\varepsilon}\cos\alpha/\kappa \\
		-i\sin\alpha\Sigma_{yx} + i\cos\alpha\Sigma_{yy} - \kappa_z\cos\alpha & \cos\alpha\Sigma_{yx} \kappa_z/\kappa + \sin\alpha\Sigma_{yy} \kappa_z/\kappa - i\sqrt{\varepsilon}\sin\alpha/\kappa
		\end{pmatrix},
\end{equation}
\end{widetext}
and $\Sigma_{ij} = 2\pi\sigma_{ij}/c$.

It can be easily checked that $\mathrm{det}\,\hat{M}_1 = 0$ only when $k_z = 0$, which makes the waves divergent at $z=\pm\infty$. So, we conclude that $\mathrm{TE}_- = \mathrm{TM}_+ = 0$, or $\mathrm{TE}_{\mathrm{I}} = \mathrm{TE}_{\mathrm{II}} = \mathrm{TE}$, $\mathrm{TM}_{\mathrm{I}} = -\mathrm{TM}_{\mathrm{II}} = \mathrm{TM}$, and arrive at
\begin{equation}
	\hat{M}_2 
	\begin{pmatrix}
		\mathrm{TE} \\
		\mathrm{TM}	
	\end{pmatrix}
	= 0.
\end{equation}
Thus, the dispersion equation reads
\begin{equation}\label{eq-disp-main}
	\mathrm{det}\,\hat{M}_2 = 0,
\end{equation}
which is fulfilled in either of the following cases:
\begin{gather}
		\label{eq-disp-ordinary-TM}
		\Sigma_0\kappa_z-i\varepsilon = 0,\; \mathrm{TE} = 0 \; \mathrm{(TM\; wave)};\\
		\text{or}\nonumber \\ 
		\label{eq-disp-TMW-1}
		\kappa_z = i\Sigma_0 \left(1- \Omega/\omega\right),\; \mathrm{TM} = 0 \; \mathrm{(TE\; wave)}.
\end{gather}

For illustrative purposes we assume the Drude model for the conductivity: $\Sigma_0 = 2\pi ie^2n_{2d}/c m\tilde{\omega}$, where $n_{2d}$ is the background sheet carrier density, $\tilde{\omega} = \omega + i/\tau_p$. 

Equation~(\ref{eq-disp-ordinary-TM}) above corresponds to ordinary TM mode with dispersion
\begin{equation}
	\label{eq-disp-ordinary-plasmon}
	\omega\tilde{\omega} = \omega_{2d}^2/\varepsilon,
\end{equation}
where $\omega_{2d}^2 = 2\pi e^2 n_0 |k_z|/m$ is the fundamental 2d plasma frequency, which we make independent of $\varepsilon$ by definition.

In turn, Eq.~(\ref{eq-disp-TMW-1}) corresponds to a two-dimensional TE thermomagnetic wave with the dispersion
\begin{equation}\label{eq-disp-TMW-2}
	\omega_{TMW} = \Omega\frac{1}{1+A^{-2}} - \frac{i}{\tau_p}\frac{1}{1+A^2},
\end{equation}
where we remind that $\Omega = -\alpha_1 c(\mathbf k,\nabla T)$, and $A = \omega_{2d}/|k_z|c$ is the retardation factor~\cite{kukushkin2003observation,muravev2020physical,zagorodnev2023two}. 

\section{Discussion}

We observe that real parts of Eqs.~(\ref{eq-disp-3DTMW}) and (\ref{eq-disp-TMW-2}) are quite similar: the 2d TMW frequency is determined by the same parameter $\Omega$, which accounts for wave non-reciprocity. Actually, Eq.~(\ref{eq-disp-TMW-2}) assumes the exact form of Eq.~(\ref{eq-disp-3DTMW}) in the fully retarded limit ($k_z \to 0, A\to\infty$, the wave is totally delocalized) if we define $\sigma_\tau^{3d} = \sigma_\tau^{2d} k_z$. Thus, we can conclude that the mode (\ref{eq-disp-TMW-2}) is the searched-for 2d thermomagnetic wave.

We checked that introduction of dissimilar dielectric permittivities $\varepsilon_1\ne\varepsilon_2$ does not lead to any qualitative changes and results only in a simple substitution $k_z\to \left(k_{z1}+k_{z2}\right)/2$ in the dispersion (\ref{eq-disp-TMW-2}), where $k_{zi} = \sqrt{k^2-\varepsilon_i k_0^2}$.

As a matter of fact, Eq.~(\ref{eq-disp-TMW-2}) is rather a condition for (and not a solution to) the 2d TMW frequency as the retardation factor $A$ depends on $k_z$ which in turn depends on $\omega$. However, in the most important limit of a strongly localized wave $k_z\simeq k\gg \sqrt{\varepsilon}k_0$, and Eq.~(\ref{eq-disp-TMW-2}) unambiguosly defines the 2d TMW frequency.
The opposite limit $k_z\simeq 0, k\simeq\sqrt{\varepsilon}k_0$ corresponds to a fully delocalized overdamped wave, described by Fal'ko and Khmelnitskii in their seminal work~\cite{falko1989if}.

We showed the 2d TMW exists and is non-reciprocal. But how well this mode is defined in terms of Q-factor? By definition,
\begin{equation}
	Q = \mathrm{Re}\,\omega/\mathrm{Im}\,\omega = \Omega\tau_p A^2,
\end{equation}
which in the limit of strong localization assumes the form:
\begin{equation}\label{eq-2dTMW-Q-factor}
	Q = -\alpha_1 \left(\mathbf{e}_\mathbf{k},\nabla T\right)\frac{2\pi}{c}\frac{e^2 n_{2d}\tau_p}{m},
\end{equation}
where $\mathbf{e}_{\mathbf{k}}$ is the unit vector in the $\mathbf{k}$ direction. We note that the 2d TMW $Q$-factor is independent of the wave vector, analogously to 2d GMW~\cite{petrov2024high}. The Q-factors of 3d TMWs and GMWs are proportional to $k^{-1}$, which stems from the different fundamental mode dispersions $\omega_{3d}\propto k^0$ vs $\omega_{2d}\propto k^{1/2}$.

Let us conduct numerical estimates of the 2d TMW $Q$-factor (\ref{eq-2dTMW-Q-factor}) and try to maximize it. Typically the Nernst-Ettingshausen coefficient $\alpha_1$ does not exceed $100\,\mu\mathrm{V}/(\mathrm{K}\cdot\mathrm{T})$~\cite{rezaei2024revealing}, carrier density is not greater than $n_{2d} = 5\cdot 10^{12}\mathrm{cm}^{-2}$, one of the lowest effective masses is achieved in bilayer graphene $m = 0.03 m_0$ ($m_0$ is the mass of a free electron), the temperature gradient is surely not greater than 100\,K/1\,$\mu$m, and the samples with best mobilities provide $\tau_p\simeq 10\,$ps. Substituting these values into Eq.~(\ref{eq-2dTMW-Q-factor}) we obtain $Q = 0.003$. Numerical solution of Eq.~(\ref{eq-disp-TMW-2}) justifies this result.

Thus, even though the absolute value of Im\,$\omega$ can be lower than the standard damping rate $1/2\tau_p$ (for $A>1$ in the long wave length limit, see Eq.~(\ref{eq-disp-TMW-2})), the $Q$-factor cannot benefit from this fact. This situation is a direct consequence of the 2d TMW localization dictated by $k_z$; in 3d case the TMW has no need in localization as it is sustained by the surrounding media.

We can suggest two possible solutions to this problem. The first option is to introduce the magnetic environment, which can add a factor of $\sqrt{\mu}$ to the $Q$-factor~\cite{petrov2024high}, where $\mu$ is the relative magnetic permittivity. Still, this will not be enough as $\mu$ does not exceed 50 in available systems~\cite{sokolov2023proximity,sokolov2024magnetic}. Another option is to search for the materials with giant \textit{dynamic} Nernst-Ettingshausen coefficient. We are unaware of any works on dynamic thermomagnetic phenomena in 2DESs despite a large amount of literature concerning the static case~\cite{rezaei2024revealing,alisultanov2014nernst,luk2011giant,zebarjadi2021nernst}.

\section{Conclusion}

In conclusion, we found a two-dimensional analog of 3d thermomagnetic waves in temperature-biased two-dimensional electron systems. These waves are isothermal, TE-polarized and non-reciprocal. The 2d TMWs arise as a result of the interaction between the dc temperature gradient and ac magnetic field in 2DESs via the Nernst-Ettingshausen effect. The Ohmic damping of 2d TMWs can be lower than that of ordinary 2d plasma waves, though 2d TMW Q-factor is extremely low. Thus, 2d TMWs can hardly be observed even in state-of-the-art 2DESs. Detailed examination of dynamic behavior of the Nernst-Ettingshausen coefficient and use of magnetic surroundings may help to overcome this issue.

\section{Acknowledgement}

The author thanks D. Svintsov for invaluable help and guidance and V. Muravev for providing reference~\cite{kopylov1978thermomagnetic} which inspired this work.

\section{Funding}

This work was supported by the Russian Science Foundation, project 23-72-01013.

\section{Conflict of interest}

The author declares no conflict of interest.

\bibliography{refs}

\providecommand{\noopsort}[1]{}\providecommand{\singleletter}[1]{#1}%
\begin{thebibliography}{24}%
\makeatletter
\providecommand \@ifxundefined [1]{%
 \@ifx{#1\undefined}
}%
\providecommand \@ifnum [1]{%
 \ifnum #1\expandafter \@firstoftwo
 \else \expandafter \@secondoftwo
 \fi
}%
\providecommand \@ifx [1]{%
 \ifx #1\expandafter \@firstoftwo
 \else \expandafter \@secondoftwo
 \fi
}%
\providecommand \natexlab [1]{#1}%
\providecommand \enquote  [1]{``#1''}%
\providecommand \bibnamefont  [1]{#1}%
\providecommand \bibfnamefont [1]{#1}%
\providecommand \citenamefont [1]{#1}%
\providecommand \href@noop [0]{\@secondoftwo}%
\providecommand \href [0]{\begingroup \@sanitize@url \@href}%
\providecommand \@href[1]{\@@startlink{#1}\@@href}%
\providecommand \@@href[1]{\endgroup#1\@@endlink}%
\providecommand \@sanitize@url [0]{\catcode `\\12\catcode `\$12\catcode `\&12\catcode `\#12\catcode `\^12\catcode `\_12\catcode `\%12\relax}%
\providecommand \@@startlink[1]{}%
\providecommand \@@endlink[0]{}%
\providecommand \url  [0]{\begingroup\@sanitize@url \@url }%
\providecommand \@url [1]{\endgroup\@href {#1}{\urlprefix }}%
\providecommand \urlprefix  [0]{URL }%
\providecommand \Eprint [0]{\href }%
\providecommand \doibase [0]{https://doi.org/}%
\providecommand \selectlanguage [0]{\@gobble}%
\providecommand \bibinfo  [0]{\@secondoftwo}%
\providecommand \bibfield  [0]{\@secondoftwo}%
\providecommand \translation [1]{[#1]}%
\providecommand \BibitemOpen [0]{}%
\providecommand \bibitemStop [0]{}%
\providecommand \bibitemNoStop [0]{.\EOS\space}%
\providecommand \EOS [0]{\spacefactor3000\relax}%
\providecommand \BibitemShut  [1]{\csname bibitem#1\endcsname}%
\let\auto@bib@innerbib\@empty
\bibitem [{\citenamefont {Jalas}\ \emph {et~al.}(2013)\citenamefont {Jalas}, \citenamefont {Petrov}, \citenamefont {Eich}, \citenamefont {Freude}, \citenamefont {Fan}, \citenamefont {Yu}, \citenamefont {Baets}, \citenamefont {Popovi{\'c}}, \citenamefont {Melloni}, \citenamefont {Joannopoulos} \emph {et~al.}}]{jalas2013and}%
  \BibitemOpen
  \bibfield  {author} {\bibinfo {author} {\bibfnamefont {D.}~\bibnamefont {Jalas}}, \bibinfo {author} {\bibfnamefont {A.}~\bibnamefont {Petrov}}, \bibinfo {author} {\bibfnamefont {M.}~\bibnamefont {Eich}}, \bibinfo {author} {\bibfnamefont {W.}~\bibnamefont {Freude}}, \bibinfo {author} {\bibfnamefont {S.}~\bibnamefont {Fan}}, \bibinfo {author} {\bibfnamefont {Z.}~\bibnamefont {Yu}}, \bibinfo {author} {\bibfnamefont {R.}~\bibnamefont {Baets}}, \bibinfo {author} {\bibfnamefont {M.}~\bibnamefont {Popovi{\'c}}}, \bibinfo {author} {\bibfnamefont {A.}~\bibnamefont {Melloni}}, \bibinfo {author} {\bibfnamefont {J.~D.}\ \bibnamefont {Joannopoulos}}, \emph {et~al.},\ }\bibfield  {title} {\bibinfo {title} {What is—and what is not—an optical isolator},\ }\href@noop {} {\bibfield  {journal} {\bibinfo  {journal} {Nature Photonics}\ }\textbf {\bibinfo {volume} {7}},\ \bibinfo {pages} {579} (\bibinfo {year} {2013})}\BibitemShut {NoStop}%
\bibitem [{\citenamefont {Yu}\ and\ \citenamefont {Fan}(2009)}]{yu2009complete}%
  \BibitemOpen
  \bibfield  {author} {\bibinfo {author} {\bibfnamefont {Z.}~\bibnamefont {Yu}}\ and\ \bibinfo {author} {\bibfnamefont {S.}~\bibnamefont {Fan}},\ }\bibfield  {title} {\bibinfo {title} {Complete optical isolation created by indirect interband photonic transitions},\ }\href@noop {} {\bibfield  {journal} {\bibinfo  {journal} {Nature photonics}\ }\textbf {\bibinfo {volume} {3}},\ \bibinfo {pages} {91} (\bibinfo {year} {2009})}\BibitemShut {NoStop}%
\bibitem [{\citenamefont {Tocci}\ \emph {et~al.}(1995)\citenamefont {Tocci}, \citenamefont {Bloemer}, \citenamefont {Scalora}, \citenamefont {Dowling},\ and\ \citenamefont {Bowden}}]{tocci1995thin}%
  \BibitemOpen
  \bibfield  {author} {\bibinfo {author} {\bibfnamefont {M.~D.}\ \bibnamefont {Tocci}}, \bibinfo {author} {\bibfnamefont {M.~J.}\ \bibnamefont {Bloemer}}, \bibinfo {author} {\bibfnamefont {M.}~\bibnamefont {Scalora}}, \bibinfo {author} {\bibfnamefont {J.~P.}\ \bibnamefont {Dowling}},\ and\ \bibinfo {author} {\bibfnamefont {C.~M.}\ \bibnamefont {Bowden}},\ }\bibfield  {title} {\bibinfo {title} {Thin-film nonlinear optical diode},\ }\href@noop {} {\bibfield  {journal} {\bibinfo  {journal} {Applied physics letters}\ }\textbf {\bibinfo {volume} {66}},\ \bibinfo {pages} {2324} (\bibinfo {year} {1995})}\BibitemShut {NoStop}%
\bibitem [{\citenamefont {Poulton}\ \emph {et~al.}(2012)\citenamefont {Poulton}, \citenamefont {Pant}, \citenamefont {Byrnes}, \citenamefont {Fan}, \citenamefont {Steel},\ and\ \citenamefont {Eggleton}}]{poulton2012design}%
  \BibitemOpen
  \bibfield  {author} {\bibinfo {author} {\bibfnamefont {C.~G.}\ \bibnamefont {Poulton}}, \bibinfo {author} {\bibfnamefont {R.}~\bibnamefont {Pant}}, \bibinfo {author} {\bibfnamefont {A.}~\bibnamefont {Byrnes}}, \bibinfo {author} {\bibfnamefont {S.}~\bibnamefont {Fan}}, \bibinfo {author} {\bibfnamefont {M.}~\bibnamefont {Steel}},\ and\ \bibinfo {author} {\bibfnamefont {B.~J.}\ \bibnamefont {Eggleton}},\ }\bibfield  {title} {\bibinfo {title} {Design for broadband on-chip isolator using stimulated brillouin scattering in dispersion-engineered chalcogenide waveguides},\ }\href@noop {} {\bibfield  {journal} {\bibinfo  {journal} {Optics express}\ }\textbf {\bibinfo {volume} {20}},\ \bibinfo {pages} {21235} (\bibinfo {year} {2012})}\BibitemShut {NoStop}%
\bibitem [{\citenamefont {Shoji}\ \emph {et~al.}(2012)\citenamefont {Shoji}, \citenamefont {Ito}, \citenamefont {Shirato},\ and\ \citenamefont {Mizumoto}}]{shoji2012mzi}%
  \BibitemOpen
  \bibfield  {author} {\bibinfo {author} {\bibfnamefont {Y.}~\bibnamefont {Shoji}}, \bibinfo {author} {\bibfnamefont {M.}~\bibnamefont {Ito}}, \bibinfo {author} {\bibfnamefont {Y.}~\bibnamefont {Shirato}},\ and\ \bibinfo {author} {\bibfnamefont {T.}~\bibnamefont {Mizumoto}},\ }\bibfield  {title} {\bibinfo {title} {Mzi optical isolator with si-wire waveguides by surface-activated direct bonding},\ }\href@noop {} {\bibfield  {journal} {\bibinfo  {journal} {Optics express}\ }\textbf {\bibinfo {volume} {20}},\ \bibinfo {pages} {18440} (\bibinfo {year} {2012})}\BibitemShut {NoStop}%
\bibitem [{\citenamefont {Petrov}\ and\ \citenamefont {Svintsov}(2024)}]{petrov2024high}%
  \BibitemOpen
  \bibfield  {author} {\bibinfo {author} {\bibfnamefont {A.~S.}\ \bibnamefont {Petrov}}\ and\ \bibinfo {author} {\bibfnamefont {D.}~\bibnamefont {Svintsov}},\ }\bibfield  {title} {\bibinfo {title} {High-frequency hall effect and transverse electric galvanomagnetic waves in current-biased 2d electron systems},\ }\href@noop {} {\bibfield  {journal} {\bibinfo  {journal} {JETP Letters}\ }\textbf {\bibinfo {volume} {119}},\ \bibinfo {pages} {800} (\bibinfo {year} {2024})}\BibitemShut {NoStop}%
\bibitem [{\citenamefont {Mikhailov}(1998)}]{mikhailov1998plasma}%
  \BibitemOpen
  \bibfield  {author} {\bibinfo {author} {\bibfnamefont {S.~A.}\ \bibnamefont {Mikhailov}},\ }\bibfield  {title} {\bibinfo {title} {Plasma instability and amplification of electromagnetic waves in low-dimensional electron systems},\ }\href@noop {} {\bibfield  {journal} {\bibinfo  {journal} {Physical Review B}\ }\textbf {\bibinfo {volume} {58}},\ \bibinfo {pages} {1517} (\bibinfo {year} {1998})}\BibitemShut {NoStop}%
\bibitem [{\citenamefont {Brown}\ and\ \citenamefont {Hoffman}(2001)}]{brown2001thermal}%
  \BibitemOpen
  \bibfield  {author} {\bibinfo {author} {\bibfnamefont {D.~C.}\ \bibnamefont {Brown}}\ and\ \bibinfo {author} {\bibfnamefont {H.~J.}\ \bibnamefont {Hoffman}},\ }\bibfield  {title} {\bibinfo {title} {Thermal, stress, and thermo-optic effects in high average power double-clad silica fiber lasers},\ }\href@noop {} {\bibfield  {journal} {\bibinfo  {journal} {IEEE Journal of quantum electronics}\ }\textbf {\bibinfo {volume} {37}},\ \bibinfo {pages} {207} (\bibinfo {year} {2001})}\BibitemShut {NoStop}%
\bibitem [{\citenamefont {Berry}(1984)}]{berry1984quantal}%
  \BibitemOpen
  \bibfield  {author} {\bibinfo {author} {\bibfnamefont {M.~V.}\ \bibnamefont {Berry}},\ }\bibfield  {title} {\bibinfo {title} {Quantal phase factors accompanying adiabatic changes},\ }\href@noop {} {\bibfield  {journal} {\bibinfo  {journal} {Proceedings of the Royal Society of London. A. Mathematical and Physical Sciences}\ }\textbf {\bibinfo {volume} {392}},\ \bibinfo {pages} {45} (\bibinfo {year} {1984})}\BibitemShut {NoStop}%
\bibitem [{\citenamefont {Xiao}\ \emph {et~al.}(2010)\citenamefont {Xiao}, \citenamefont {Chang},\ and\ \citenamefont {Niu}}]{xiao2010berry}%
  \BibitemOpen
  \bibfield  {author} {\bibinfo {author} {\bibfnamefont {D.}~\bibnamefont {Xiao}}, \bibinfo {author} {\bibfnamefont {M.-C.}\ \bibnamefont {Chang}},\ and\ \bibinfo {author} {\bibfnamefont {Q.}~\bibnamefont {Niu}},\ }\bibfield  {title} {\bibinfo {title} {Berry phase effects on electronic properties},\ }\href@noop {} {\bibfield  {journal} {\bibinfo  {journal} {Reviews of modern physics}\ }\textbf {\bibinfo {volume} {82}},\ \bibinfo {pages} {1959} (\bibinfo {year} {2010})}\BibitemShut {NoStop}%
\bibitem [{\citenamefont {Petrov}(2021)}]{petrov2021plasmonic}%
  \BibitemOpen
  \bibfield  {author} {\bibinfo {author} {\bibfnamefont {A.~S.}\ \bibnamefont {Petrov}},\ }\bibfield  {title} {\bibinfo {title} {Plasmonic excitation for a tunable transmitter without magnetic field immune to backscattering},\ }\href@noop {} {\bibfield  {journal} {\bibinfo  {journal} {Physical Review B}\ }\textbf {\bibinfo {volume} {104}},\ \bibinfo {pages} {L241407} (\bibinfo {year} {2021})}\BibitemShut {NoStop}%
\bibitem [{\citenamefont {Gurevich}(1963)}]{gurevich1963thermomagnetic}%
  \BibitemOpen
  \bibfield  {author} {\bibinfo {author} {\bibfnamefont {L.}~\bibnamefont {Gurevich}},\ }\bibfield  {title} {\bibinfo {title} {Thermomagnetic waves and excitation of a magnetic field in a nonequilibrium plasma},\ }\href@noop {} {\bibfield  {journal} {\bibinfo  {journal} {Zh. Eksperim. i Teor. Fiz.}\ }\textbf {\bibinfo {volume} {44}} (\bibinfo {year} {1963})}\BibitemShut {NoStop}%
\bibitem [{\citenamefont {Gurevich}\ and\ \citenamefont {Gel'Mont}(1967)}]{gurevich1967nonlinear}%
  \BibitemOpen
  \bibfield  {author} {\bibinfo {author} {\bibfnamefont {L.}~\bibnamefont {Gurevich}}\ and\ \bibinfo {author} {\bibfnamefont {B.}~\bibnamefont {Gel'Mont}},\ }\bibfield  {title} {\bibinfo {title} {Nonlinear theory of thermomagnetic waves},\ }\href@noop {} {\bibfield  {journal} {\bibinfo  {journal} {Soviet Journal of Experimental and Theoretical Physics}\ }\textbf {\bibinfo {volume} {24}},\ \bibinfo {pages} {124} (\bibinfo {year} {1967})}\BibitemShut {NoStop}%
\bibitem [{\citenamefont {Kopylov}(1978)}]{kopylov1978thermomagnetic}%
  \BibitemOpen
  \bibfield  {author} {\bibinfo {author} {\bibfnamefont {V.}~\bibnamefont {Kopylov}},\ }\bibfield  {title} {\bibinfo {title} {Thermomagnetic waves in bismuth},\ }\href@noop {} {\bibfield  {journal} {\bibinfo  {journal} {JETP Lett}\ }\textbf {\bibinfo {volume} {28}} (\bibinfo {year} {1978})}\BibitemShut {NoStop}%
\bibitem [{\citenamefont {Kukushkin}\ \emph {et~al.}(2003)\citenamefont {Kukushkin}, \citenamefont {Smet}, \citenamefont {Mikhailov}, \citenamefont {Kulakovskii}, \citenamefont {Von~Klitzing},\ and\ \citenamefont {Wegscheider}}]{kukushkin2003observation}%
  \BibitemOpen
  \bibfield  {author} {\bibinfo {author} {\bibfnamefont {I.}~\bibnamefont {Kukushkin}}, \bibinfo {author} {\bibfnamefont {J.}~\bibnamefont {Smet}}, \bibinfo {author} {\bibfnamefont {S.~A.}\ \bibnamefont {Mikhailov}}, \bibinfo {author} {\bibfnamefont {D.}~\bibnamefont {Kulakovskii}}, \bibinfo {author} {\bibfnamefont {K.}~\bibnamefont {Von~Klitzing}},\ and\ \bibinfo {author} {\bibfnamefont {W.}~\bibnamefont {Wegscheider}},\ }\bibfield  {title} {\bibinfo {title} {Observation of retardation effects in the spectrum of two-dimensional plasmons},\ }\href@noop {} {\bibfield  {journal} {\bibinfo  {journal} {Physical review letters}\ }\textbf {\bibinfo {volume} {90}},\ \bibinfo {pages} {156801} (\bibinfo {year} {2003})}\BibitemShut {NoStop}%
\bibitem [{\citenamefont {Muravev}\ \emph {et~al.}(2020)\citenamefont {Muravev}, \citenamefont {Gusikhin}, \citenamefont {Zarezin}, \citenamefont {Zabolotnykh}, \citenamefont {Volkov},\ and\ \citenamefont {Kukushkin}}]{muravev2020physical}%
  \BibitemOpen
  \bibfield  {author} {\bibinfo {author} {\bibfnamefont {V.}~\bibnamefont {Muravev}}, \bibinfo {author} {\bibfnamefont {P.}~\bibnamefont {Gusikhin}}, \bibinfo {author} {\bibfnamefont {A.}~\bibnamefont {Zarezin}}, \bibinfo {author} {\bibfnamefont {A.}~\bibnamefont {Zabolotnykh}}, \bibinfo {author} {\bibfnamefont {V.}~\bibnamefont {Volkov}},\ and\ \bibinfo {author} {\bibfnamefont {I.}~\bibnamefont {Kukushkin}},\ }\bibfield  {title} {\bibinfo {title} {Physical origin of relativistic plasmons in a two-dimensional electron system},\ }\href@noop {} {\bibfield  {journal} {\bibinfo  {journal} {Physical Review B}\ }\textbf {\bibinfo {volume} {102}},\ \bibinfo {pages} {081301} (\bibinfo {year} {2020})}\BibitemShut {NoStop}%
\bibitem [{\citenamefont {Zagorodnev}\ \emph {et~al.}(2023)\citenamefont {Zagorodnev}, \citenamefont {Zabolotnykh}, \citenamefont {Rodionov},\ and\ \citenamefont {Volkov}}]{zagorodnev2023two}%
  \BibitemOpen
  \bibfield  {author} {\bibinfo {author} {\bibfnamefont {I.~V.}\ \bibnamefont {Zagorodnev}}, \bibinfo {author} {\bibfnamefont {A.~A.}\ \bibnamefont {Zabolotnykh}}, \bibinfo {author} {\bibfnamefont {D.~A.}\ \bibnamefont {Rodionov}},\ and\ \bibinfo {author} {\bibfnamefont {V.~A.}\ \bibnamefont {Volkov}},\ }\bibfield  {title} {\bibinfo {title} {Two-dimensional plasmons in laterally confined 2d electron systems},\ }\href@noop {} {\bibfield  {journal} {\bibinfo  {journal} {Nanomaterials}\ }\textbf {\bibinfo {volume} {13}},\ \bibinfo {pages} {975} (\bibinfo {year} {2023})}\BibitemShut {NoStop}%
\bibitem [{\citenamefont {Falko}\ and\ \citenamefont {Khmelnitskii}(1989)}]{falko1989if}%
  \BibitemOpen
  \bibfield  {author} {\bibinfo {author} {\bibfnamefont {V.}~\bibnamefont {Falko}}\ and\ \bibinfo {author} {\bibfnamefont {D.}~\bibnamefont {Khmelnitskii}},\ }\bibfield  {title} {\bibinfo {title} {What if a film conductivity exceeds the speed of light},\ }\href@noop {} {\bibfield  {journal} {\bibinfo  {journal} {Zh. Eksp. Teor. Fiz}\ }\textbf {\bibinfo {volume} {95}},\ \bibinfo {pages} {847} (\bibinfo {year} {1989})}\BibitemShut {NoStop}%
\bibitem [{\citenamefont {Rezaei}\ and\ \citenamefont {Schindler}(2024)}]{rezaei2024revealing}%
  \BibitemOpen
  \bibfield  {author} {\bibinfo {author} {\bibfnamefont {S.~E.}\ \bibnamefont {Rezaei}}\ and\ \bibinfo {author} {\bibfnamefont {P.}~\bibnamefont {Schindler}},\ }\bibfield  {title} {\bibinfo {title} {Revealing large room-temperature nernst coefficients in 2d materials by first-principles modeling},\ }\href@noop {} {\bibfield  {journal} {\bibinfo  {journal} {Nanoscale}\ }\textbf {\bibinfo {volume} {16}},\ \bibinfo {pages} {6142} (\bibinfo {year} {2024})}\BibitemShut {NoStop}%
\bibitem [{\citenamefont {Sokolov}\ \emph {et~al.}(2023)\citenamefont {Sokolov}, \citenamefont {Averyanov}, \citenamefont {Parfenov}, \citenamefont {Taldenkov}, \citenamefont {Rybin}, \citenamefont {Tokmachev},\ and\ \citenamefont {Storchak}}]{sokolov2023proximity}%
  \BibitemOpen
  \bibfield  {author} {\bibinfo {author} {\bibfnamefont {I.~S.}\ \bibnamefont {Sokolov}}, \bibinfo {author} {\bibfnamefont {D.~V.}\ \bibnamefont {Averyanov}}, \bibinfo {author} {\bibfnamefont {O.~E.}\ \bibnamefont {Parfenov}}, \bibinfo {author} {\bibfnamefont {A.~N.}\ \bibnamefont {Taldenkov}}, \bibinfo {author} {\bibfnamefont {M.~G.}\ \bibnamefont {Rybin}}, \bibinfo {author} {\bibfnamefont {A.~M.}\ \bibnamefont {Tokmachev}},\ and\ \bibinfo {author} {\bibfnamefont {V.~G.}\ \bibnamefont {Storchak}},\ }\bibfield  {title} {\bibinfo {title} {Proximity coupling of graphene to a submonolayer 2d magnet},\ }\href@noop {} {\bibfield  {journal} {\bibinfo  {journal} {Small}\ ,\ \bibinfo {pages} {2301295}} (\bibinfo {year} {2023})}\BibitemShut {NoStop}%
\bibitem [{\citenamefont {Sokolov}\ \emph {et~al.}(2024)\citenamefont {Sokolov}, \citenamefont {Averyanov}, \citenamefont {Parfenov}, \citenamefont {Taldenkov}, \citenamefont {Tokmachev},\ and\ \citenamefont {Storchak}}]{sokolov2024magnetic}%
  \BibitemOpen
  \bibfield  {author} {\bibinfo {author} {\bibfnamefont {I.~S.}\ \bibnamefont {Sokolov}}, \bibinfo {author} {\bibfnamefont {D.~V.}\ \bibnamefont {Averyanov}}, \bibinfo {author} {\bibfnamefont {O.~E.}\ \bibnamefont {Parfenov}}, \bibinfo {author} {\bibfnamefont {A.~N.}\ \bibnamefont {Taldenkov}}, \bibinfo {author} {\bibfnamefont {A.~M.}\ \bibnamefont {Tokmachev}},\ and\ \bibinfo {author} {\bibfnamefont {V.~G.}\ \bibnamefont {Storchak}},\ }\bibfield  {title} {\bibinfo {title} {Magnetic heterostructure of graphene with a submonolayer magnet on silicon},\ }\href@noop {} {\bibfield  {journal} {\bibinfo  {journal} {Carbon}\ }\textbf {\bibinfo {volume} {218}},\ \bibinfo {pages} {118769} (\bibinfo {year} {2024})}\BibitemShut {NoStop}%
\bibitem [{\citenamefont {Alisultanov}(2014)}]{alisultanov2014nernst}%
  \BibitemOpen
  \bibfield  {author} {\bibinfo {author} {\bibfnamefont {Z.~Z.}\ \bibnamefont {Alisultanov}},\ }\bibfield  {title} {\bibinfo {title} {Nernst-ettingshausen effect in graphene},\ }\href@noop {} {\bibfield  {journal} {\bibinfo  {journal} {JETP letters}\ }\textbf {\bibinfo {volume} {99}},\ \bibinfo {pages} {702} (\bibinfo {year} {2014})}\BibitemShut {NoStop}%
\bibitem [{\citenamefont {Luk’yanchuk}\ \emph {et~al.}(2011)\citenamefont {Luk’yanchuk}, \citenamefont {Varlamov},\ and\ \citenamefont {Kavokin}}]{luk2011giant}%
  \BibitemOpen
  \bibfield  {author} {\bibinfo {author} {\bibfnamefont {I.~A.}\ \bibnamefont {Luk’yanchuk}}, \bibinfo {author} {\bibfnamefont {A.~A.}\ \bibnamefont {Varlamov}},\ and\ \bibinfo {author} {\bibfnamefont {A.~V.}\ \bibnamefont {Kavokin}},\ }\bibfield  {title} {\bibinfo {title} {Giant nernst-ettingshausen oscillations in semiclassically strong magnetic fields},\ }\href@noop {} {\bibfield  {journal} {\bibinfo  {journal} {Physical Review Letters}\ }\textbf {\bibinfo {volume} {107}},\ \bibinfo {pages} {016601} (\bibinfo {year} {2011})}\BibitemShut {NoStop}%
\bibitem [{\citenamefont {Zebarjadi}\ \emph {et~al.}(2021)\citenamefont {Zebarjadi}, \citenamefont {Rezaei}, \citenamefont {Akhanda},\ and\ \citenamefont {Esfarjani}}]{zebarjadi2021nernst}%
  \BibitemOpen
  \bibfield  {author} {\bibinfo {author} {\bibfnamefont {M.}~\bibnamefont {Zebarjadi}}, \bibinfo {author} {\bibfnamefont {S.~E.}\ \bibnamefont {Rezaei}}, \bibinfo {author} {\bibfnamefont {M.~S.}\ \bibnamefont {Akhanda}},\ and\ \bibinfo {author} {\bibfnamefont {K.}~\bibnamefont {Esfarjani}},\ }\bibfield  {title} {\bibinfo {title} {Nernst coefficient within relaxation time approximation},\ }\href@noop {} {\bibfield  {journal} {\bibinfo  {journal} {Physical Review B}\ }\textbf {\bibinfo {volume} {103}},\ \bibinfo {pages} {144404} (\bibinfo {year} {2021})}\BibitemShut {NoStop}%
\end{thebibliography}%

\end{document}